\documentclass[aps,prb,reprint,floatfix,amsmath,amssymb,longbibliography]{revtex4-2}
\usepackage{bm}
\usepackage{graphicx}
\graphicspath{{figures/}{./}}
\usepackage[hidelinks]{hyperref}
\newcommand{\Tr}{\operatorname{Tr}}
\newcommand{\erfc}{\operatorname{erfc}}
\newcommand{\dd}{\mathrm{d}}
\makeatletter
\AtBeginDocument{\immediate\write\@auxout{\string\citation{apsrev42Control}}}
\makeatother
\begin{document}
\title{Accelerating real-time hybrid-functional calculations with learned ACE and symmetry-based FFT reuse}
\author{Tomohito Otobe}
\email{otobe.tomohito@qst.go.jp}
\affiliation{Kansai Institute for Photon Science,
National Institutes for Quantum Science and Technology, Umemidai 8-1-7, Kizugawa, Kyoto, Japan 619-0215}
\date{}
\begin{abstract}
Real-time hybrid-functional calculations require repeated reconstruction of nonlocal exchange, whose cost grows rapidly with Brillouin-zone sampling. We develop two methods to improve the efficiency of time evolution and spatial processing. The first learns the time evolution of adaptively compressed exchange (ACE) factors and predicts their updates, reducing the frequency of explicit reconstruction. The second uses crystal symmetry to reuse exchange fast Fourier transforms (FFTs) among equivalent real-space points. This symmetry-based FFT reuse (SBFR) reduces the number of FFTs while retaining the full $\bm k$-point sum. The methods can be used with a variety of hybrid functionals, and we use HSE06 in this work. Relative to an ACE reference using full-row mesh FFTs, learned prediction provides more than fivefold acceleration, SBFR more than fourteenfold, and their combination more than sixty-threefold. The combined method shows efficient parallel scaling. After learning, propagation takes less than 0.5 s per step. The reduced computational cost allows us to study nonlinear and nonequilibrium electron dynamics with hybrid functionals and to examine the role of excitonic effects in semiconductors and insulators.
\end{abstract}
\maketitle

\section{Introduction}
Real-time time-dependent density-functional theory (RT-TDDFT) describes electronic response to external fields \cite{Runge1984,Yabana1996RealTime,Bertsch2000Dielectric,Meng2008Dynamics}. Nonlinear and nonequilibrium phenomena in solids \cite{Xu2024Nonequilibrium} include laser-induced optical breakdown \cite{Otobe2008Breakdown}, coherent phonon generation \cite{Shinohara2010CoherentPhonon}, pump--probe response \cite{Sato2014PumpProbe}, the dynamical Franz--Keldysh effect \cite{Otobe2016DFKE}, and high-harmonic generation \cite{Ghimire2011CrystalHHG,TancogneDejean2017HHG}. Their accurate description requires reliable band gaps and electron--hole interactions, which semilocal approximations often inadequately capture \cite{Reining2002}. Hybrid functionals such as HSE06 improve this description through nonlocal exchange \cite{Heyd2003,Krukau2006,Heyd2005Gaps} and have been applied to nonlinear excitation, optical spectra, and excitonic response \cite{Sato2015HybridDynamics,Ketolainen2020,Pemmaraju2019Excitons,Xu2026AllElectron}.

Hybrid functionals such as HSE and PBE0 are substantially more expensive than semilocal functionals such as PBE because of nonlocal exchange \cite{Lin2016}. This cost is a major obstacle in real-time calculations. Nonlocal exchange couples each target orbital to occupied orbitals throughout the Brillouin zone. As these orbitals evolve, exchange must be updated over thousands of time steps \cite{Jia2019}, and finer $\bm k$-point sampling further increases the work. This cost makes systematic calculations over laser intensity, frequency, and polarization difficult, even for a small unit cell. Reducing the exchange cost is therefore essential for studying nonlinear dynamics routinely with hybrid functionals.

Adaptively compressed exchange (ACE) reduces repeated exchange-operator application costs \cite{Lin2016}; combined with parallel-transport integration, it enables efficient hybrid RT-TDDFT, including finite-temperature simulations \cite{Jia2019,Liu2025}. Density fitting, orbital locality, and low-rank representations further reduce exchange work \cite{Lu2015THC,Hu2017,Ko2023SeA,Li2023RTACEISDF}, with Fourier convolution \cite{Wu2022KPointACEISDF} and tensor hypercontraction \cite{Rettig2023THC} addressing $\bm k$-point sampling. Despite these advances, explicit exchange evaluation during ACE reconstruction remains costly. In this work, we make hybrid-functional propagation more practical by reducing both the reconstruction frequency in time and the FFT work in space.

The first method accelerates time evolution by learning the changes in ACE factors and reducing the reconstruction frequency. Previous learning approaches model molecular time-dependent Hartree--Fock potentials \cite{Bhat2024TDHFLearning} or electron-density propagators \cite{Shah2026TimePropagators}. We fit the coefficients to gauge-aligned factor increments during propagation and use them to predict the factors between explicit corrections. The orbitals, density, and semilocal potentials are updated at every time step.

The second method improves spatial processing by reducing the number of fast Fourier transforms (FFTs) in each reconstruction. We achieve this through symmetry-based FFT reuse (SBFR). Representative-orbital transformations \cite{VASPISYM}, symmetry-reduced exchange-matrix construction with localized resolution of identity \cite{Cao2025Symmetry}, and symmetry-adapted density fitting for many-body calculations \cite{Yeh2024SymmetryISDF} provide related precedents. We reuse mesh FFTs of symmetry-equivalent density-matrix rows directly on the real-space grid, without fitting orbital products or reducing the exchange sum over $\bm k'$. Spatial-pair distribution parallelizes the remaining FFTs beyond the number of independent rows.

We implement both methods in SALMON \cite{Noda2019} and demonstrate them with HSE06. Section~\ref{sec:theory} presents the methods, Sec.~\ref{sec:measures} defines the validation measures, and Sec.~\ref{sec:results} reports silicon benchmarks. Section~\ref{sec:conclusions} summarizes the findings and outlook.

\section{Theory and methods}
\label{sec:theory}

\subsection{Screened exchange in real time}
Atomic units are used unless stated otherwise. HSE replaces a fraction of short-range Perdew--Burke--Ernzerhof (PBE) exchange by Hartree--Fock exchange, retaining PBE correlation and long-range exchange \cite{Perdew1996PBE}. Its exchange-correlation energy is
\begin{align}
E_{xc}^{\mathrm{HSE}}={}&aE_x^{\mathrm{HF,SR}}(\omega_{\mathrm{SR}})
 +(1-a)E_x^{\mathrm{PBE,SR}}(\omega_{\mathrm{SR}})\nonumber\\
 &+E_x^{\mathrm{PBE,LR}}(\omega_{\mathrm{SR}})+E_c^{\mathrm{PBE}},
\end{align}
where HSE06 uses $a=1/4$ and $\omega_{\mathrm{SR}}=0.11$ \cite{Krukau2006}. For each spin channel, define the one-particle density matrix
\begin{equation}
\gamma_\sigma(\bm r,\bm r';t)=\sum_{n\bm k}w_{\bm k}f_{n\bm k\sigma}
\psi_{n\bm k\sigma}(\bm r,t)\psi_{n\bm k\sigma}^{*}(\bm r',t).
\end{equation}
Here $f_{n\bm k\sigma}$ is the occupation per spin, and $w_{\bm k}$ is the normalized Brillouin-zone weight. The screened exchange action is
\begin{equation}
\begin{aligned}
{}[V_{x,\sigma}^{\mathrm{SR}}\psi_{n\bm k\sigma}](\bm r,t)={}&-\sum_{\bm R}\int_{\Omega}\dd^3\bm r'\,
\mathcal K_0(\bm r-\bm r'-\bm R)\\
&\times\gamma_\sigma(\bm r,\bm r'+\bm R;t)\psi_{n\bm k\sigma}(\bm r'+\bm R,t).
\end{aligned}
\label{eq:fock}
\end{equation}
Here $\Omega$ denotes the real-space simulation box and, when used as a scalar, its volume. The vectors $\bm R$ are lattice translations and $\mathcal K_0(\bm r)=\erfc(\omega_{\mathrm{SR}}|\bm r|)/|\bm r|$. Translated orbitals carry their Bloch phases.
For an occupied orbital $\psi_{n'\bm k'\sigma}$ and target orbital $\psi_{n\bm k\sigma}$, the pair density is $\psi_{n'\bm k'\sigma}^{*}\psi_{n\bm k\sigma}$. Explicit exchange evaluates the corresponding convolution and multiplies its result by $\psi_{n'\bm k'\sigma}$. The source sum is weighted by the occupations.

For occupied-space reconstruction, direct exchange requires convolutions over occupied-band pairs and source and target wave vectors. We use a uniform real-space grid with $N_r$ points; the resulting costs are summarized in Table~\ref{tab:cost}.

Real-time propagation follows the self-consistent time-dependent Kohn--Sham equation \cite{Runge1984,Sugino1999Dynamics,Castro2004Propagators},
\begin{equation}
i\partial_t\psi_{n\bm k}=
\left[h_{\mathrm{sl}}[n(t),\bm A(t)]+aV_x^{\mathrm{SR}}[\gamma(t)]\right]\psi_{n\bm k},
\end{equation}
where $h_{\mathrm{sl}}$ includes kinetic, ionic, Hartree, and the appropriate semilocal HSE terms.

\subsection{ACE construction and application}
To reduce the cost of repeated exchange applications during propagation, we construct ACE separately at each target $\bm k$ point from its $N_{\mathrm{occ}}$ occupied orbitals. Wave-vector and spin labels are suppressed in this subsection. The matrix $\Phi$ collects these orbitals, and $W=V_x^{\mathrm{SR}}\Phi$ contains their explicit exchange images. The overlap matrix is defined by spatial integration,
\[
M_{ij}=\int_{\Omega}\phi_i^*(\bm r)W_j(\bm r)\,\dd^3r.
\]
In operator notation this gives
\begin{equation}
M=\Phi^\dagger W,\qquad -M=LL^\dagger,
\label{eq:chol}
\end{equation}
with $L$ lower triangular. For a nonsingular negative-definite overlap matrix, define
\begin{equation}
\Xi=WL^{-\dagger},\qquad V_x^{\mathrm{ACE}}=-\Xi\Xi^\dagger.
\label{eq:ace}
\end{equation}
Here $L^{-\dagger}=(L^\dagger)^{-1}$ is evaluated by triangular solves. Since $M^{-1}=-L^{-\dagger}L^{-1}$,
\begin{equation}
V_x^{\mathrm{ACE}}=WM^{-1}W^\dagger,
\qquad V_x^{\mathrm{ACE}}\Phi=W.
\end{equation}
Thus ACE reproduces explicit exchange on the construction subspace.

For any target $\phi$, the ACE action is
\begin{equation}
V_x^{\mathrm{ACE}}\phi=-\sum_{\mu=1}^{N_{\mathrm{occ}}}
\xi_\mu(\bm r)\int_{\Omega}\xi_\mu^*(\bm r')\phi(\bm r')\,\dd^3r',
\end{equation}
where $\xi_\mu(\bm r)$ is the $\mu$th column of $\Xi$.
ACE replaces pairwise FFTs with two matrix products for overlaps and linear combinations. The factors are reused between explicit reconstructions.

ACE agrees closely with direct screened exchange in the laser-driven Si benchmark (Appendix~\ref{app:ace-direct}).

\subsection{Temporal acceleration with learned ACE}
\label{sec:learned-method}
ACE makes exchange application inexpensive, but reconstructing its factors at every time step remains costly. To reduce this work, we predict the factors from their recent evolution and retain periodic explicit corrections. The density and semilocal potentials are updated at every step. Related history-based methods reduce self-consistency work in polarizable molecular dynamics \cite{Kolafa2004} and expensive force evaluations in ACE-based Born--Oppenheimer molecular dynamics \cite{Mandal2019MTS}.

Unitary column rotations leave the ACE operator unchanged but can introduce artificial variations in its factor history. Before fitting or prediction, we align successive explicit factors by a complex extension of orthogonal Procrustes alignment \cite{Schonemann1966}; see Appendix~\ref{app:prediction-details}.

The aligned explicit factor at endpoint $m$ is $\Xi_m$, with endpoints separated by $K\Delta t$. Within an interval, $s=\ell/K$ for $\ell=1,\ldots,K-1$. To learn the factor increments, we retain three explicit factors and form
\begin{equation}
D_1=\Xi_m-\Xi_{m-1},\qquad D_2=\Xi_{m-1}-\Xi_{m-2}.
\end{equation}
For factor functions, we define the real-valued inner product
\[
(A,B)=\Re\sum_\mu\int_{\Omega} A_\mu^*(\bm r)B_\mu(\bm r)\,\dd^3r.
\]
Here $\mu$ labels factor columns and $\|A\|_F^2=(A,A)$.
For a later explicit endpoint, the target increment is $Y=\Xi_{m+1}-\Xi_m$. The two real coefficients $c=(c_1,c_2)$ are determined by cumulative regularized least squares,
\begin{equation}
\min_{c_1,c_2}\sum_m\|Y_m-c_1D_{1,m}-c_2D_{2,m}\|_F^2
+\lambda(c_1^2+c_2^2).
\end{equation}
To stabilize the fit for nearly dependent increments, we set $\lambda=10^{-8}\sum_m(\|D_{1,m}\|_F^2+\|D_{2,m}\|_F^2)$.
The fit determines two real coefficients for each target-$\bm k$ history.

The learned predictor adds a correction to linear extrapolation, giving
\begin{equation}
\begin{aligned}
\widehat{\Xi}_{m,\ell}={}&\Xi_m+\bigl[s+s^2(c_1-1)\bigr]D_1\\
&+s^2c_2D_2.
\end{aligned}
\end{equation}
It reproduces the learned endpoint increment at $s=1$ and reduces to linear extrapolation when $c=(1,0)$. At $\ell=K$, explicit exchange supplies the next factor and updates the fit. For either predictor, the exchange operator is constructed as $-\widehat{\Xi}\widehat{\Xi}^\dagger$, preserving its algebraic sign and Hermiticity.

During warmup, ACE is reconstructed at every time step, while training increments are accumulated at intervals of $K$ steps. We set $N_{\mathrm{hi}}=3$ history factors and $N_{\mathrm{teach}}=8$ training increments. These determine the warmup length automatically, giving 80 steps at $K=8$ (Appendix~\ref{app:prediction-details}).

\subsection{Spatial acceleration with SBFR}
\label{sec:symmetry-method}
Crystal symmetry makes many real-space contributions to exchange equivalent. To avoid repeating the corresponding FFTs, we first perform the occupied-band sum and express exchange through the density matrix resolved by $\bm k'$. For a mesh with reference point $\bm k_0$, we use the phase convention
\begin{align}
v_{n\bm k'}(\bm r)
 &=e^{-i\bm k_0\cdot\bm r}\psi_{n\bm k'}(\bm r)\nonumber\\
 &=e^{i(\bm k'-\bm k_0)\cdot\bm r}u_{n\bm k'}(\bm r),\\
\bar\gamma_{\bm k'}(\bm r,\bm r')
 &=\sum_{n\in\mathrm{occ}}v_{n\bm k'}(\bm r)v_{n\bm k'}^{*}(\bm r').
\end{align}
Here $\psi_{n\bm k'}=e^{i\bm k'\cdot\bm r}u_{n\bm k'}$ is a Bloch orbital and $\bm k_0$ is the first point of the full uniform mesh. Unit occupations per spin and equal weights $1/N_k$ are assumed; spin and time labels are suppressed. For each coordinate pair, the mesh transforms are
\begin{align}
\widehat\gamma_{\bm R}(\bm r,\bm r')
 &=\sum_{\bm k'\in\mathcal M_k}e^{-i(\bm k'-\bm k_0)\cdot\bm R}
 \bar\gamma_{\bm k'}(\bm r,\bm r'),\label{eq:dense-forward}\\
B_{\bm k}(\bm r,\bm r')
 &=\sum_{\bm R\in\mathcal B}e^{i(\bm k-\bm k_0)\cdot\bm R}
 \mathcal K_B(\bm r-\bm r'-\bm R)
 \widehat\gamma_{\bm R}(\bm r,\bm r'),\label{eq:dense-backward}
\end{align}
where $\mathcal M_k$ is the complete $N_k$-point mesh, $\mathcal B$ is its dual set of lattice translations in the Born--von Karman supercell, and $\mathcal K_B$ is the screened kernel periodic on that supercell. These transforms act along the $\bm k'$ and $\bm R$ indices, with both spatial coordinates held fixed. The exchange action is
\begin{equation}
\begin{aligned}
{}[V_x^{\mathrm{SR}}u_{n\bm k}](\bm r)
={}&-\frac{e^{-i(\bm k-\bm k_0)\cdot\bm r}}{N_k}\\
&\times\int_\Omega B_{\bm k}(\bm r,\bm r')v_{n\bm k}(\bm r')\,\dd^3r'.
\end{aligned}
\label{eq:dense-action}
\end{equation}
Appendix~\ref{app:mesh-exchange} derives these expressions and their discrete normalization.

The real-space representation is particularly useful for this procedure because symmetry-related grid points are connected by coordinate permutations and phase factors. We can therefore reuse equivalent density-matrix rows and distribute the remaining FFTs over spatial pairs independently of orbital ownership.

Crystal symmetry reduces orbital propagation to representative $\bm k$ points. At each $\bm k'$, we copy the orbitals from its corresponding representative point by permuting the real-space grid according to the symmetry operation and multiplying by the associated Bloch phase factors. We form $\bar\gamma_{\bm k'}$ from these orbitals, averaging the projectors when several symmetry operations reach the same $\bm k'$. SBFR additionally reuses FFT results among symmetry-equivalent real-space rows: Eqs.~\eqref{eq:dense-forward} and \eqref{eq:dense-backward} are evaluated only at inequivalent first coordinates $\bm r$, while $\bm r'$ spans the full cell. The remaining rows of $B_{\bm k}$ are restored by symmetry for contraction in Eq.~\eqref{eq:dense-action}. Thus representative-point reduction decreases the number of independently propagated orbitals, whereas SBFR decreases the number of mesh FFTs; each FFT and the exchange sum retain all $N_k$ points.

Symmetry reconstruction transforms the wave vector and both spatial coordinates together, including the Bloch phases associated with cell wrapping (Appendix~\ref{app:mesh-exchange}). Retained operations must preserve the occupied projector, discrete kernel, and external field. For the Si tests, we use 32 space-group operations preserving the $z$-directed field, including fractional translations.

Choosing $\bm r_s$ from the representative first-coordinate points reduces the number of coordinate pairs requiring mesh FFTs from
\begin{equation}
N_r^2\ \longrightarrow\ N_aN_r,
\qquad N_{k'}=N_k\ \text{in both routes},
\label{eq:dense-work-reduction}
\end{equation}
where $N_a$ is the number of symmetry-inequivalent real-space grid points. For the $12^3$ grid used here, $N_r=1728$ and $N_a=75$, reducing the number of pairs requiring FFTs by a factor of 23. We distribute the $N_aN_r$ pairs over MPI ranks in batches. Each task performs a full $N_k$-point transform, symmetry restoration, and partial target contraction, avoiding storage of the full two-coordinate density matrix. The SBFR Off reference uses the same representative orbitals and mesh transforms but evaluates all first-coordinate rows. Timings include projector generation, FFTs, and target contraction; detailed costs are given in Appendix~\ref{app:cost-details}.

\subsection{Summary of computational costs}
Table~\ref{tab:cost} summarizes the computational work. Here $N_{\mathrm{occ}}$ is the occupied-orbital count per $\bm k$ point, $N_r$ the grid size, $N_a$ the number of symmetry-inequivalent grid points, and $P_k$ the MPI rank count used for $\bm k$-point and spatial-pair distributions.
\begin{table}[t]
\caption{Leading computational work, excluding MPI padding and communication. Costs are totals unless marked per rank. ACE combinations give post-learning averages; the mesh-FFT rows cover transforms and kernel multiplication only.}
\label{tab:cost}
\small
\setlength{\tabcolsep}{3pt}
\begin{ruledtabular}
\begin{tabular}{ll}
Method or operation & Computational work\\
\hline
\shortstack[l]{Direct exchange\\(spatial FFT)} & $O(N_k^2N_{\mathrm{occ}}^2N_r\log N_r)$\\
\shortstack[l]{Full-row mesh FFTs\\(Off reference)} & $O(N_r^2N_k\log N_k)$\\
\shortstack[l]{SBFR\\mesh FFTs} & $O(N_aN_rN_k\log N_k)$\\
\shortstack[l]{Distributed mesh FFTs\\(per rank)} & $O(N_aN_rN_k\log N_k/P_k)$\\
ACE & $C_{\mathrm{apply}}+C_{\mathrm{build}}^{\mathrm{global}}$\\
Learned ACE & $C_{\mathrm{apply}}+C_{\mathrm{predict}}+C_{\mathrm{build}}^{\mathrm{global}}/K$\\
SBFR + ACE & $C_{\mathrm{apply}}+C_{\mathrm{build}}^{\mathrm{sym}}$\\
\shortstack[l]{SBFR\\+ learned ACE} & $C_{\mathrm{apply}}+C_{\mathrm{predict}}+C_{\mathrm{build}}^{\mathrm{sym}}/K$\\
\end{tabular}
\end{ruledtabular}
\end{table}
The common costs are $C_{\mathrm{apply}}=O(N_kN_{\mathrm{occ}}^2N_r)$ for ACE application and $C_{\mathrm{predict}}=O(N_kN_{\mathrm{occ}}N_r)$ for prediction at fixed history length. The reconstruction costs $C_{\mathrm{build}}^{\mathrm{global}}$ and $C_{\mathrm{build}}^{\mathrm{sym}}$ refer to conventional exchange and SBFR, including factor construction. In the performance benchmarks, $C_{\mathrm{build}}^{\mathrm{global}}$ uses the full-row mesh-FFT reference. Symmetry reduces mesh-FFT work by $N_r/N_a$, while prediction reduces reconstruction frequency by $K$. The per-rank estimate assumes balanced spatial-pair tasks with $P_k\lesssim N_aN_r$. At fixed cell and $P_k\propto N_k$, the mesh-FFT contribution per rank grows as $O(\log N_k)$. Communication, padding, and the other reconstruction costs are discussed in Appendix~\ref{app:cost-details}.

\section{Validation and performance measures}
\label{sec:measures}
We compare learned prediction with conventional ACE and SBFR On with SBFR Off, keeping all other calculation settings fixed.

The absorbed energy per cell is evaluated from the matter-current density as
\begin{equation}
\Delta E_{\mathrm{abs}}(t)=-\Omega\int_0^t
\bm E_{\mathrm{tot}}(t')\cdot\bm J(t')\,\dd t'.
\end{equation}
Here $\bm E_{\mathrm{tot}}$ is the total electric field and $\bm J$ is the matter-current density output by SALMON.

The HHG spectrum is evaluated as
\begin{equation}
S(\omega)=\omega^2\left|\int_0^T w(t)J_z(t)e^{i\omega t}\,\dd t\right|^2,
\end{equation}
using a common window $w(t)=\sin^2(\pi t/T)$ over the trajectory duration $T$.

We quantify the current error over the complete trajectory by
\begin{equation}
\epsilon_J=\left[
\frac{\displaystyle\int_0^T\!\left|J_z(t)-J_z^{\mathrm{ref}}(t)\right|^2\dd t}
{\displaystyle\int_0^T\!\left|J_z^{\mathrm{ref}}(t)\right|^2\dd t}
\right]^{1/2}.
\end{equation}
The energy error is the relative difference of final absorbed energies.

For transverse impulse-response calculations \cite{Bertsch2000Dielectric}, we obtain the conductivity and dielectric function from
\begin{align}
\sigma_{zz}(\omega)&=\frac{1}{\mathcal I_z}\int_0^T
w_{\mathrm{imp}}(t)J_z(t)e^{i\omega t}\,\dd t,\\
\epsilon_{zz}(\omega)&=1+\frac{4\pi i}{\omega}\sigma_{zz}(\omega),
\end{align}
where $\mathcal I_z$ is the applied electric-field impulse amplitude in the SALMON convention and $w_{\mathrm{imp}}(t)=1-3(t/T)^2+2(t/T)^3$. We evaluate the transform by $\Delta t\sum_j$ using the recorded currents at $t_j=j\Delta t$.

Performance is reported as propagation time per step, with learning and post-learning intervals separated. Step timings average the maximum duration across MPI ranks at each step and exclude checkpoint and timer-record output. Each configuration is timed once.

\section{Results}
\label{sec:results}

We use an eight-atom conventional Si cell with 32 valence electrons. We set the cell side length to 10.26 a.u., the real-space grid to $12^3$ points, and the ground-state orbital count to 32. We use a norm-conserving Si pseudopotential generated in-house with the Perdew--Zunger local-density parametrization \cite{Perdew1981PZ}, treating four valence electrons per atom. We set uniform $\bm k$ meshes shifted by half a mesh spacing in each reciprocal direction. Real-time calculations propagate the 16 occupied orbitals using fourth-order Taylor propagation with predictor--corrector updates and $\Delta t=0.16$ a.u. The screened kernel is untruncated.

\subsection{Numerical assessment of predictive ACE}
\label{sec:learned-validation}
Reducing the reconstruction frequency is useful only if prediction preserves the laser-driven response. To determine the correction interval, we compare learned ACE with conventional ACE on a $4^3$ $\bm k$ mesh over a 15-fs trajectory. We set the laser photon energy to 1.55 eV and the peak intensity to $10^{12}\,\mathrm{W\,cm^{-2}}$. We use a $z$-polarized vector potential with a $\cos^2$ envelope of 10-fs full support. We vary the explicit correction interval over $K=2$, 4, 8, 12, and 16.

\begin{figure}[tb]
\centering
\includegraphics[width=\columnwidth]{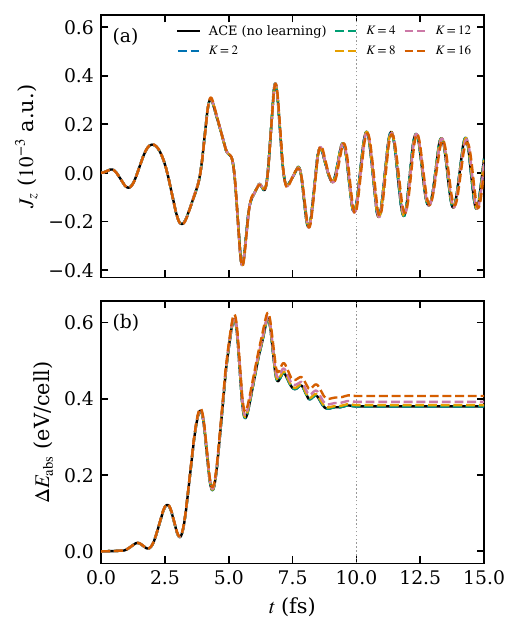}
\caption{Laser-driven (a) matter-current density and (b) absorbed energy per conventional Si cell for HSE06 with $N_k=4^3$. Learned ACE with correction intervals $K=2$, 4, 8, 12, and 16 is compared with ACE without learning. The dotted line marks the pulse end.}
\label{fig:learned-waveforms}
\end{figure}

\begin{figure}[t]
\centering
\includegraphics[width=\columnwidth]{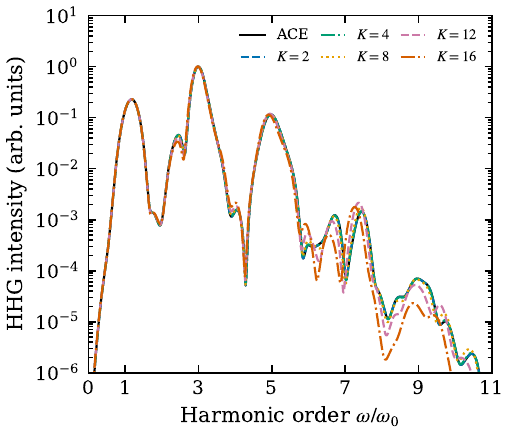}
\caption{HHG spectra from conventional ACE and learned ACE with correction intervals $K=2$, 4, 8, 12, and 16 for the laser parameters in Sec.~\ref{sec:learned-validation}. The horizontal axis is the harmonic order, with $\hbar\omega_0=1.55$ eV. All spectra use the same 15-fs window and are normalized by the maximum conventional-ACE intensity over the displayed range.}
\label{fig:learned-hhg}
\end{figure}

Figure~\ref{fig:learned-waveforms} compares the induced current and absorbed energy; the conventional-ACE reference absorbs 0.38 eV per cell. At $K=4$ and $K=8$, prediction also reproduces the conventional-ACE high-harmonic spectrum through the ninth harmonic (Fig.~\ref{fig:learned-hhg}).

\begin{figure}[tb]
\centering
\includegraphics[width=\columnwidth]{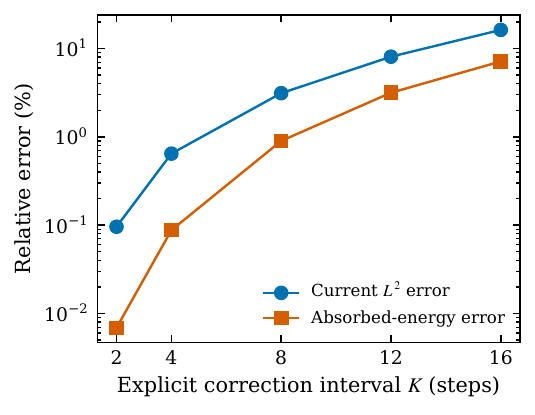}
\caption{Relative errors of the current and absorbed energy versus explicit correction interval $K$ for the calculations in Fig.~\ref{fig:learned-waveforms}. Current errors use the full-record $L^2$ norm. Absorbed-energy errors use the final field-current integral. The reference is ACE without temporal learning. Lines guide the eye.}
\label{fig:learned-accuracy}
\end{figure}

Figure~\ref{fig:learned-accuracy} quantifies the current and absorbed-energy errors. At $K=8$, these errors are 3.1\% and 0.90\%, whereas they increase to 16\% and 7.1\% at $K=16$. The lower reconstruction cost at $K=16$ thus comes at the expense of response accuracy. We therefore set $K=8$ for the following calculations. At this interval, learning reduces the current error from 11\% for linear extrapolation to 3.1\% at essentially the same cost (Appendix~\ref{app:linear-prediction}).

\subsection{SBFR performance}
\label{sec:symmetry-results}
To separate the gains from temporal prediction and spatial FFT reuse, we compare the two methods individually and in combination.

\label{sec:fugaku-dense-combined}
We conduct the benchmarks on Fugaku. Each node has one Fujitsu A64FX CPU with 48 compute cores and 32 GiB of high-bandwidth HBM2 memory, providing a peak memory bandwidth of 1024 GB/s. Nodes are connected by Tofu Interconnect D in a six-dimensional mesh/torus topology \cite{FugakuSystem}.

Exchange tests confirm that SBFR preserves numerical precision across shifted and unshifted meshes, fractional translations, MPI layouts, and FFT array layouts. For timing, we compare conventional ACE, learned ACE, and learned ACE with SBFR at $N_k=4^3$, using full-row mesh FFTs for both SBFR Off cases. We assign the 12 representative $\bm k$ points to 12 MPI ranks on three nodes and set 12 OpenMP threads per rank, giving one representative point per rank. We retain crystal symmetry and all $4^3$ source $\bm k'$ points in every case. We apply a $z$-directed impulse of $10^{-4}$ a.u. and propagate for 160 steps (0.62 fs). We average the step times over steps 1--80 during learning and 81--160 after learning, with 70 prediction steps and 10 explicit corrections in the latter interval.

\begin{table*}[tb]
\caption{Fugaku performance for $N_k=4^3$. SBFR Off and SBFR On use full-row and SBFR mesh FFTs, respectively, with identical orbital symmetry reduction. Speedups are relative to conventional ACE.}
\label{tab:fugaku-dense-combined}
\begin{ruledtabular}
\begin{tabular}{lcrrrr}
Method & SBFR & During learning (s/step) & Speedup & After learning (s/step) & Speedup\\
\hline
Conventional ACE & Off & 36.909 & 1.00 & 36.889 & 1.00\\
Learned ACE & Off & 36.680 & 1.01 & 6.978 & 5.29\\
Learned ACE & On & 2.539 & 14.54 & 0.582 & 63.39\\
\end{tabular}
\end{ruledtabular}
\end{table*}
Table~\ref{tab:fugaku-dense-combined} shows the two contributions to the acceleration. During learning, explicit exchange is evaluated at every step, and SBFR alone reduces the step time by a factor of 15. After learning, prediction reduces the reconstruction frequency and gives a 5.3-fold acceleration without SBFR. Combining the two methods gives a 63-fold acceleration relative to conventional ACE.

SBFR reduces explicit correction steps from 55 to 3.8 s, while prediction steps remain near 0.13 s. It therefore accelerates the costly steps that temporal prediction cannot eliminate.

Both scaling tests use this combined method with the same impulse and timing intervals, propagating symmetry-inequivalent representative $\bm k$ points while retaining the full mesh in the exchange sum. We set four MPI ranks per node and 12 OpenMP threads per rank for both tests.

\subsection{Strong scaling}
\begin{figure}[tb]
\centering
\includegraphics[width=\columnwidth]{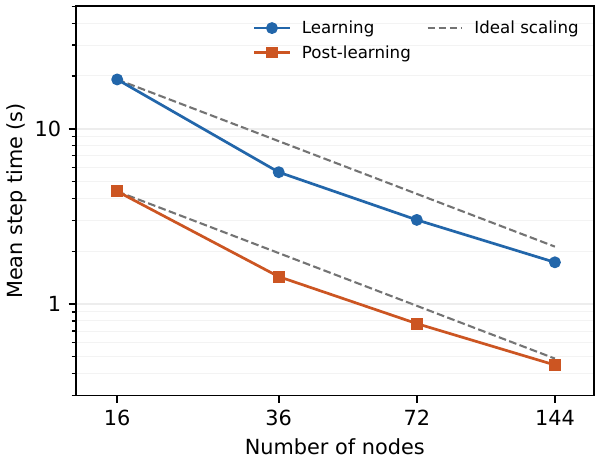}
\caption{Fugaku strong scaling for a conventional Si cell at $N_k=16^3$, using learned ACE and SBFR with spatial-pair distribution. Dashed lines indicate inverse-node-count references anchored to the two 16-node timings.}
\label{fig:pairfft-strong-scaling}
\end{figure}

To examine whether a fixed problem can be solved efficiently with more MPI resources, we first study strong scaling. We set $N_k=16^3$ with 576 representative points and use 16, 36, 72, and 144 nodes, corresponding to 64, 144, 288, and 576 MPI ranks, respectively. Figure~\ref{fig:pairfft-strong-scaling} shows that, from 16 to 144 nodes, the learning time decreases from 19 to 1.7 s/step, while the post-learning time decreases from 4.4 to 0.45 s/step, giving speedups of 11 and 9.8, respectively, for a ninefold increase in node count. Orbitals are distributed over representative $\bm k$ points, whereas exchange FFTs are distributed over spatial-grid pairs $(\bm r,\bm r')$. Increasing the number of ranks reduces the number of spatial pairs held by each rank. The resulting smaller FFT working set may fit more effectively in cache and explain the slightly superlinear speedups.

\subsection{Weak scaling}
We next examine whether the step time can be maintained as the $\bm k$ mesh grows. For this weak-scaling test, we increase the MPI resources to keep one symmetry-inequivalent representative $\bm k$ point per rank. For $N_k=4^3$, $8^3$, $16^3$, and $32^3$, symmetry gives 12, 80, 576, and 4352 representative points on 3, 20, 144, and 1088 nodes, respectively.

\begin{figure*}[tb]
\centering
\includegraphics[width=\textwidth]{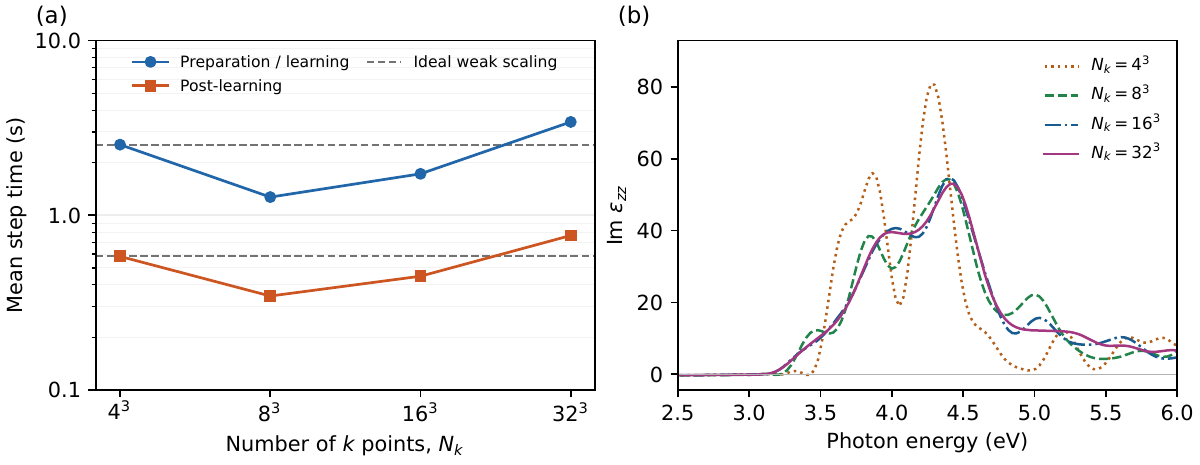}
\caption{Fugaku weak scaling and optical response for a conventional Si cell. (a) Mean time per step during learning and after learning versus the full mesh size $N_k$, with one representative $\bm k$ point per MPI rank. Dashed lines indicate constant-time references at $N_k=4^3$. (b) Imaginary dielectric response for $N_k=4^3$, $8^3$, $16^3$, and $32^3$ from 20-fs impulse-response calculations.}
\label{fig:pairfft-size-scaling}
\label{fig:kmesh-optical}
\end{figure*}

Figure~\ref{fig:pairfft-size-scaling}(a) compares the learning and post-learning step times with constant-time reference lines. The post-learning cost remains below 0.80 s/step over the full series. It decreases from 0.58 to 0.35 s/step between $4^3$ and $8^3$, then rises to 0.45 and 0.77 s/step at $16^3$ and $32^3$. The initial decrease may likewise reflect improved cache utilization, as discussed for strong scaling. The learning curve follows the same trend. From $8^3$ to $32^3$, a 64-fold increase in the full $\bm k$ mesh increases the post-learning time by a factor of 2.2.

We can understand this increase by separating prediction steps from explicit corrections. Prediction steps cost 0.13--0.15 s across the series, whereas explicit corrections increase from 1.9 s at $8^3$ to 5.1 s at $32^3$. Although the number of propagated representative points per rank is fixed, each pair FFT retains the growing full source mesh. Reconstruction every eight steps limits this contribution to the average cost (Appendix~\ref{app:weak-scaling-timings}).

To determine the sampling needed for the linear response, we compare the dielectric spectra in Fig.~\ref{fig:kmesh-optical}(b). The structure of the coarse-mesh spectrum becomes smoother as the number of points increases. The $N_k=16^3$ and $32^3$ spectra are close to each other, indicating that $\bm k$-point convergence is nearly reached on this real-space grid.

\subsection{Laser-driven response with $k$-point sampling}
\begin{figure*}[tb]
\centering
\includegraphics[width=\textwidth]{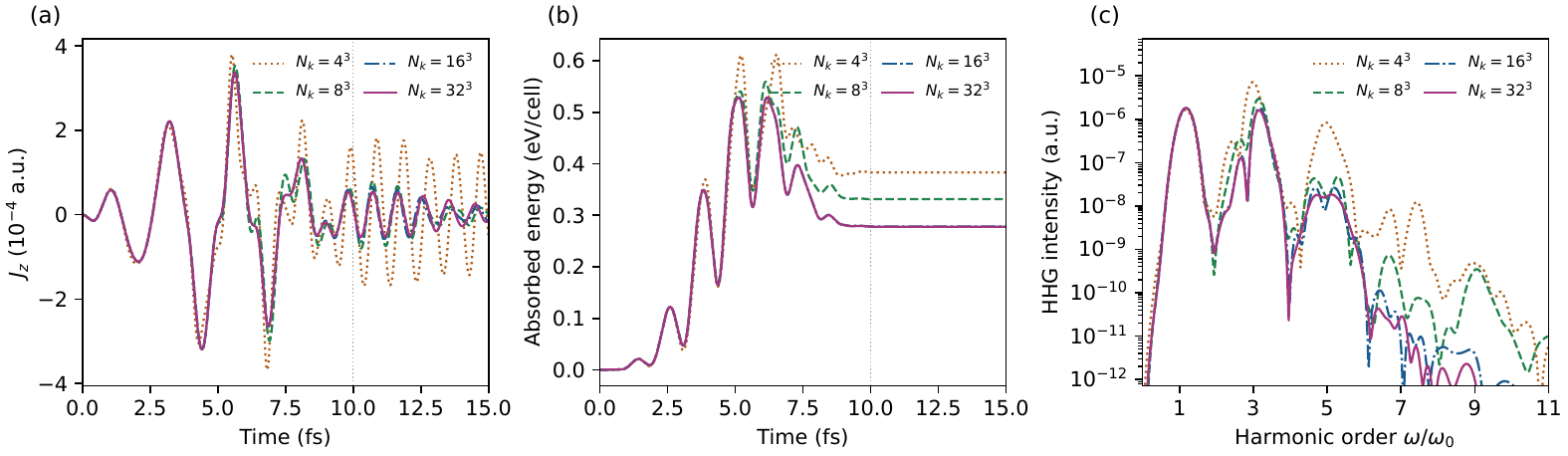}
\caption{Laser-driven Si response for $N_k=4^3$, $8^3$, $16^3$, and $32^3$: (a) matter-current density, (b) absorbed energy per cell, and (c) HHG spectra. The laser parameters are those of Sec.~\ref{sec:learned-validation}. Dotted vertical lines mark the pulse end.}
\label{fig:kmesh-laser}
\end{figure*}

To determine whether the same sampling also describes nonlinear excitation, we compare the laser-driven responses at $N_k=4^3$, $8^3$, $16^3$, and $32^3$ in Fig.~\ref{fig:kmesh-laser}, using the pulse of Sec.~\ref{sec:learned-validation} and the same learning schedule. The $N_k=32^3$ calculation over 15 fs required $3.2\times10^3$ s (about 54 min), including initialization, on 1088 Fugaku nodes. The pronounced post-pulse current oscillations of the coarsest mesh are suppressed with finer sampling. The final absorbed energies are 0.38, 0.33, 0.28, and 0.28 eV per cell, respectively; the change from $16^3$ to $32^3$ is only 0.085\%. The current waveforms also nearly converge at $N_k=16^3$, with a full-record relative difference of 6.1\% from $32^3$. Above the seventh harmonic, differences between the two finest meshes are still visible, although the qualitative spectral features are unchanged. Under these conditions, $16^3$ sampling describes the main response. Finer sampling is needed for quantitative comparison of the higher harmonics.

\section{Conclusions}
\label{sec:conclusions}
We have accelerated real-time hybrid-functional calculations in two directions: time evolution with learned ACE and spatial processing with SBFR. The two reductions can be combined because they act on the frequency and cost of exchange reconstruction, respectively.

For silicon, we found a 5.3-fold acceleration from learned prediction after learning and a 15-fold acceleration from SBFR during learning, using full-row mesh-FFT ACE as the reference. Combining the two methods gives a 63-fold acceleration after learning. For the tested laser pulse, $K=8$ balances reconstruction frequency and response accuracy. Spatial-pair parallelism enables strong scaling of the remaining FFT work. The post-learning cost reaches 0.45 s/step at $N_k=16^3$ on 144 Fugaku nodes. It remains below 0.80 s/step through $N_k=32^3$ with one representative point per rank. With these step times, we can perform repeated HSE06 calculations to examine $\bm k$-point convergence and the dependence of electron dynamics on laser parameters.

The methods also apply to PBE0 and other hybrid functionals. Their improved description of the band gap and electron--hole interaction is important for nonlinear electron dynamics in semiconductors and insulators. The reduced cost makes it practical to examine how excitonic effects appear under strong laser excitation. By combining the present methods with SALMON's multiscale framework \cite{Yabana2012Multiscale,Noda2019}, we can further treat electron dynamics and light propagation simultaneously and approach experimental irradiation conditions more accurately.

\begin{acknowledgments}
This work was supported by the MEXT Quantum Leap Flagship Program (MEXT Q-LEAP), Grant No.~JPMXS0118067246; JSPS KAKENHI, Grant No.~24K23024; and the MEXT Next-Generation Computational Science Global Reach \& Innovation Program (GRIP), Japan Grant Number JPMXP1920260201. This work used computational resources of the supercomputer Fugaku provided by the RIKEN Center for Computational Science (Project ID: hp260497).
The research ideas, computational approach, and analysis procedures were specified by the author. OpenAI Codex (GPT-6.1 Sol) was used under the author's direction to assist with code development, numerical postprocessing, and manuscript preparation. The author reviewed and verified the resulting code, analyses, and manuscript.
\end{acknowledgments}

\section*{Data availability}
The implementation described in this article is planned for inclusion in a future release of SALMON. The input files will be made publicly available on GitHub. The numerical data supporting the findings are available from the author upon reasonable request.

\appendix

\section{Detailed timings for weak scaling}
\label{app:weak-scaling-timings}
Table~\ref{tab:pairfft-size-scaling} gives the step times underlying Fig.~\ref{fig:pairfft-size-scaling}(a), including the prediction and explicit-correction contributions.

\begin{table*}[tb]
\caption{Step times for the size-scaling series in Fig.~\ref{fig:pairfft-size-scaling}(a). All times are in seconds per step. The learning and post-learning columns average steps 1--80 and 81--160, respectively. Prediction-only and explicit-correction times are averaged separately within the post-learning interval.}
\label{tab:pairfft-size-scaling}
\begin{ruledtabular}
\begin{tabular}{crrrrrr}
$\bm k$ mesh & Nodes & MPI ranks & During learning & Post-learning & Prediction only & Explicit correction\\
\hline
$4^3$ & 3 & 12 & 2.539 & 0.582 & 0.127 & 3.763\\
$8^3$ & 20 & 80 & 1.272 & 0.345 & 0.129 & 1.858\\
$16^3$ & 144 & 576 & 1.728 & 0.448 & 0.148 & 2.546\\
$32^3$ & 1088 & 4352 & 3.424 & 0.767 & 0.150 & 5.086\\
\end{tabular}
\end{ruledtabular}
\end{table*}
\section{Density-matrix formulation of mesh exchange}
\label{app:mesh-exchange}
Write a Bloch orbital as $\psi_{n\bm k}(\bm r)=e^{i\bm k\cdot\bm r}u_{n\bm k}(\bm r)$, where $u_{n\bm k}$ is cell periodic. The periodic pair density and its spatial Fourier transform are
\begin{align}
\rho_{n'n}^{\bm k'\bm k}(\bm r')
 &=u_{n'\bm k'}^{*}(\bm r')u_{n\bm k}(\bm r'),\\
\widetilde\rho_{n'n}^{\bm k'\bm k}(\bm G)
 &=\int_\Omega e^{-i\bm G\cdot\bm r'}
 \rho_{n'n}^{\bm k'\bm k}(\bm r')\,\dd^3r'.
\end{align}
Here $\bm G$ is a reciprocal lattice vector. The usual exchange convolution multiplies this transform by $\widetilde{\mathcal K}_{\mathrm{SR}}(\bm G+\bm q)$ and transforms back to real space, where $\bm q=\bm k-\bm k'$ is the momentum transfer. The screened multiplier is
\begin{equation}
\widetilde{\mathcal K}_{\mathrm{SR}}(\bm Q)
=\frac{4\pi}{Q^2}\left[1-e^{-Q^2/(4\omega_{\mathrm{SR}}^2)}\right],
\label{eq:dense-screened-multiplier}
\end{equation}
with limit $\pi/\omega_{\mathrm{SR}}^2$ at $Q=0$. For an unscreened hybrid such as PBE0, the multiplier becomes $4\pi/Q^2$, and its $Q=0$ Coulomb singularity requires a separate treatment; the factor prediction and SBFR construction otherwise remain applicable. Here $\bm Q=\bm G+\bm q$.

To change the order of evaluation, use Bloch periodicity in Eq.~\eqref{eq:fock},
\begin{equation}
\begin{aligned}
&\psi_{n'\bm k'}^{*}(\bm r'+\bm R)
 \psi_{n\bm k}(\bm r'+\bm R)\\
&\qquad=e^{i(\bm k-\bm k')\cdot\bm R}
 \psi_{n'\bm k'}^{*}(\bm r')\psi_{n\bm k}(\bm r').
\end{aligned}
\end{equation}
The occupied-orbital sum can then be performed before the sums over $\bm k'$ and lattice translations. The density matrix contains products of the same occupied orbital at two positions, whereas the pair density contains an occupied orbital and the target orbital at the same position. Their relation is
\begin{equation}
\begin{aligned}
&\sum_{n'\in\mathrm{occ}}u_{n'\bm k'}(\bm r)
 \rho_{n'n}^{\bm k'\bm k}(\bm r')\\
&\quad=\left[\sum_{n'\in\mathrm{occ}}u_{n'\bm k'}(\bm r)
 u_{n'\bm k'}^{*}(\bm r')\right]u_{n\bm k}(\bm r').
\end{aligned}
\label{eq:pair-projector-reordering}
\end{equation}
The target orbital thus factors out of the occupied-band sum.

We use the uniform mesh
\begin{equation}
\bm k=\bm k_0+\sum_{\alpha=1}^{3}\frac{j_\alpha}{M_\alpha}\bm b_\alpha,
\qquad 0\leq j_\alpha<M_\alpha,
\label{eq:dense-mesh-origin}
\end{equation}
where $\bm b_\alpha$ are reciprocal lattice vectors and $N_k=M_1M_2M_3$. The origin $\bm k_0$ may describe an unshifted or shifted mesh. Its dual translations are $\bm R=\sum_\alpha l_\alpha\bm a_\alpha$, with $0\leq l_\alpha<M_\alpha$ and $\bm a_\alpha\cdot\bm b_\beta=2\pi\delta_{\alpha\beta}$. These translations form the set $\mathcal B$ in the Born--von Karman supercell. The sum over $\bm k'$ runs over the full mesh $\mathcal M_k$.

With the rephased orbitals and occupied projector defined in Sec.~\ref{sec:symmetry-method}, the relation to the density matrix in Eq.~(2) is
\begin{align}
\bar\gamma_{\bm k'}(\bm r,\bm r')
 &=e^{-i\bm k_0\cdot(\bm r-\bm r')}
 \sum_{n\in\mathrm{occ}}\psi_{n\bm k'}(\bm r)\psi_{n\bm k'}^*(\bm r'),\\
\gamma(\bm r,\bm r')
 &=\frac{e^{i\bm k_0\cdot(\bm r-\bm r')}}{N_k}
 \sum_{\bm k'}\bar\gamma_{\bm k'}(\bm r,\bm r').
\end{align}
Thus $\bar\gamma_{\bm k'}$ is the density matrix resolved by $\bm k'$, with a common phase removed to accommodate a shifted mesh. Let $\mathcal K_B$ be the screened kernel made periodic on the Born--von Karman supercell. Folding the lattice translations into $\mathcal B$ gives
\begin{equation}
\begin{aligned}
{}[V_x^{\mathrm{SR}}u_{n\bm k}](\bm r)
={}&-\frac{e^{-i(\bm k-\bm k_0)\cdot\bm r}}{N_k}
 \int_\Omega\dd^3r'\,v_{n\bm k}(\bm r')\\
&\times\sum_{\bm R\in\mathcal B}\mathcal K_B(\bm r-\bm r'-\bm R)
 e^{i(\bm k-\bm k_0)\cdot\bm R}\\
&\times\sum_{\bm k'\in\mathcal M_k}
 e^{-i(\bm k'-\bm k_0)\cdot\bm R}\bar\gamma_{\bm k'}(\bm r,\bm r').
\end{aligned}
\label{eq:dense-untransformed}
\end{equation}
The innermost sum is a discrete Fourier transform because
\[
(\bm k'-\bm k_0)\cdot\bm R
=2\pi\sum_\alpha j_\alpha l_\alpha/M_\alpha.
\]
For each fixed coordinate pair, Eqs.~\eqref{eq:dense-forward} and \eqref{eq:dense-backward} evaluate the two mesh transforms with kernel multiplication between them. The remaining spatial integration gives Eq.~\eqref{eq:dense-action}, equivalent to Eq.~\eqref{eq:dense-untransformed} with the full $\bm k'$ sum.

The spatial Fourier transform is performed when preparing $\mathcal K_B$ from Eq.~\eqref{eq:dense-screened-multiplier}. On the $N_rN_k$-point supercell grid, the implementation stores the quadrature-weighted kernel
\begin{equation}
\mathcal K(\bm d)=\frac{1}{N_rN_k}\operatorname{Re}
\sum_{\bm Q\in\mathcal Q_{\mathrm{BvK}}}
\widetilde{\mathcal K}_{\mathrm{SR}}(\bm Q)e^{i\bm Q\cdot\bm d},
\label{eq:dense-kernel-normalization}
\end{equation}
where $\mathcal Q_{\mathrm{BvK}}$ is the supercell reciprocal grid. The real part imposes the discrete even-kernel convention at Nyquist modes. Using this stored kernel in Eq.~\eqref{eq:dense-backward} includes the spatial quadrature in the final contraction. The mesh FFTs use the unnormalized FFTW convention, leaving the exchange weight $1/N_k$ explicit in Eq.~\eqref{eq:dense-action}.

The reconstruction must transform the wave vector and both spatial coordinates together. For a space-group operation $s=\{S|\bm t\}$, define cell coordinates by
\begin{equation}
S\bm r_s+\bm t=\bm r+\bm L_s(\bm r_s),
\label{eq:dense-wrapped-coordinate}
\end{equation}
where $S$ is the orthogonal matrix for the rotational part (including improper rotations), $\bm t$ is the translation vector, $\bm L_s$ is a lattice translation, and $\bm k_s$ is the mesh representative of $S^{-1}\bm k$ modulo a reciprocal lattice vector. Covariance of the occupied projector gives
\begin{equation}
\begin{aligned}
B_{\bm k}(\bm r,\bm r')={}&
 e^{i\bm k\cdot[\bm L_s(\bm r_s')-\bm L_s(\bm r_s)]}\\
&\times e^{i\bm k_0\cdot[(\bm r_s-\bm r)-(\bm r_s'-\bm r')]}
 B_{\bm k_s}(\bm r_s,\bm r_s').
\end{aligned}
\label{eq:dense-row-symmetry}
\end{equation}
The first phase accounts for cell wrapping. The second comes from rephasing the Bloch orbitals and vanishes when $\bm k_0=0$. No individual-band phase choice is required because common orbital phases cancel in the projector.

\section{Factor alignment and learning schedule}
\label{app:prediction-details}
For an unaligned factor $\Xi^{\mathrm{raw}}$ and the latest accepted factor $\Xi^{\mathrm{ref}}$, the column rotation solves
\begin{equation}
Q_\star=\mathop{\mathrm{arg\,min}}_{Q^\dagger Q=I}
\|\Xi^{\mathrm{raw}}Q-\Xi^{\mathrm{ref}}\|_F^2.
\end{equation}
The factor norms are invariant under $Q$, so minimizing this distance is equivalent to maximizing $\Re\Tr[Q^\dagger M]$, where $M=(\Xi^{\mathrm{raw}})^\dagger\Xi^{\mathrm{ref}}$. If $M=U\Sigma V^\dagger$ is its singular-value decomposition, $Q_\star=UV^\dagger$ and the aligned factor is $\Xi=\Xi^{\mathrm{raw}}Q_\star$. Since $Q_\star Q_\star^\dagger=I$, this transformation leaves $-\Xi\Xi^\dagger$ unchanged. Each explicit factor is aligned to the already aligned reference, so the stored factors and their increments share a common column gauge.

The cumulative least-squares fit requires only a real symmetric $2\times2$ matrix and a two-component vector per target $\bm k$ history. For accepted training samples, we accumulate
\begin{align}
G_{ab}&=\sum_m(D_{a,m},D_{b,m}),\\
b_a&=\sum_m(D_{a,m},Y_m),\qquad a,b=1,2.
\end{align}
Differentiating the regularized objective gives
\begin{equation}
\begin{pmatrix}G_{11}+\lambda&G_{12}\\G_{12}&G_{22}+\lambda\end{pmatrix}
\begin{pmatrix}c_1\\c_2\end{pmatrix}
=\begin{pmatrix}b_1\\b_2\end{pmatrix},
\end{equation}
where $\lambda=10^{-8}(G_{11}+G_{22})$. Only explicit endpoints supply training targets $Y_m$. The quadratic interpolation in $s$ retains the latest increment as its initial slope and approaches the fitted increment at the next correction endpoint.

The predicted ACE factor is held fixed during the internal predictor--corrector stages of each time step.

Starting from one explicit factor at the initial time, $N_{\mathrm{hi}}-1$ correction intervals are needed to populate the factor history, followed by $N_{\mathrm{teach}}$ intervals to collect the training samples. The default number of warmup time steps is therefore
\begin{equation}
N_{\mathrm{warm}}=(N_{\mathrm{hi}}-1+N_{\mathrm{teach}})K,
\end{equation}
where $N_{\mathrm{hi}}$ is the number of stored factors, $N_{\mathrm{teach}}$ is the minimum number of explicit target increments $Y_m$ required for fitting before prediction starts, and $K$ is the number of time steps between explicit corrections. Each training sample comprises $(D_{1,m},D_{2,m},Y_m)$. For the history and training counts used here, $N_{\mathrm{warm}}=10K$. Linear extrapolation uses the same prediction-start schedule.

\section{Detailed computational and memory costs}
\label{app:cost-details}
The mesh-FFT estimates in Table~\ref{tab:cost} include transforms and kernel multiplication. Projector generation, symmetry restoration, target contraction, and ACE factor construction contribute to the full reconstruction cost $C_{\mathrm{build}}^{\mathrm{sym}}$. ACE application and prediction use the symmetry-inequivalent target points, whereas each reconstructed exchange action retains the full source mesh.

At fixed $N_k$, grid spacing, and density, $N_r$ and $N_{\mathrm{occ}}$ grow in proportion to system size $N$. For fixed $N_a/N_r$, the SBFR mesh-FFT work is $O(N^2)$, but orbital contractions and ACE factor construction retain an $O(N^3)$ cost, including $O(N_kN_{\mathrm{occ}}^3)$ dense orbital-space operations. At fixed $K$, learned prediction reduces reconstruction frequency without changing its system-size exponent. Alignment and coefficient fitting are included in the explicit-update cost for learned ACE.

For the CPU distribution over spatial coordinate pairs, a pass assigns at most $c$ pairs $(\bm r_a,\bm r')$ to each of the $P_k$ ranks. The number of passes and allocated pair slots are
\begin{equation}
q_c=\left\lceil\frac{N_aN_r}{P_kc}\right\rceil,
\qquad N_{\mathrm{pair}}^{\mathrm{pad}}=P_kcq_c.
\end{equation}
Each active rank executes forward and backward mesh FFTs for $c$ arrays of length $N_k$ per pass. Ranks without a valid pair skip the FFTs but participate in the transposes. The executed-work upper bounds are
\begin{align}
W_{\mathrm{FFT}}^{\mathrm{rank}}&=O(q_c cN_k\log N_k),\\
W_{\mathrm{FFT}}^{\mathrm{all}}&=O(N_{\mathrm{pair}}^{\mathrm{pad}}N_k\log N_k).
\end{align}
Ignoring batch padding, balanced assignment of the independent coordinate pairs gives the maximum local FFT work
\begin{equation}
W_{\mathrm{FFT}}^{\mathrm{rank,max}}
=O\!\left[\max\!\left(1,\frac{N_aN_r}{P_k}\right)N_k\log N_k\right].
\end{equation}
Thus, spatial-pair parallelism reduces the local FFT work up to $P_k\sim N_aN_r$, where each active rank handles one pair.

The batch size $c$ balances the coordinate pairs across ranks subject to a workspace bound. For the present $12^3$ grid, $N_a=75$ gives 129600 independent pairs. Negligible padding recovers the useful-work estimate $O(N_aN_rN_k\log N_k)$.

Each pass uses two fixed-count all-to-all transposes. If $n_{k,\max}$ is the largest number of full-mesh points assigned to one rank, the per-rank payload over one reconstruction is
\begin{equation}
V_{\mathrm{transpose}}^{\mathrm{rank}}
=O(q_cP_kc\,n_{k,\max})
\end{equation}
in complex elements, excluding collective latency and protocol overhead. For balanced full-mesh ownership, $n_{k,\max}=O(N_k/P_k)$ and the payload is $O(q_ccN_k)$ per rank. In the balanced regime $q_cc=O(N_aN_r/P_k)$, the local FFT work and payload are $O(N_aN_rN_k\log N_k/P_k)$ and $O(N_aN_rN_k/P_k)$, respectively. Communication time also depends on the collective algorithm and network mapping.

ACE factors are distributed over local real-space rows. Their overlaps require spatial reductions, while local matrix products and grid operations use OpenMP. For complex double precision, the three stored histories require
\begin{equation}
B_{\mathrm{hist}}=3\times16\,N_{r,\mathrm{local}}
N_{\mathrm{occ}}N_{k,\mathrm{local}}\quad\text{bytes}.
\end{equation}
Peak memory additionally includes the current and predicted factors, orbitals, and workspaces.

\begin{figure}[!t]
\centering
\includegraphics[width=\columnwidth]{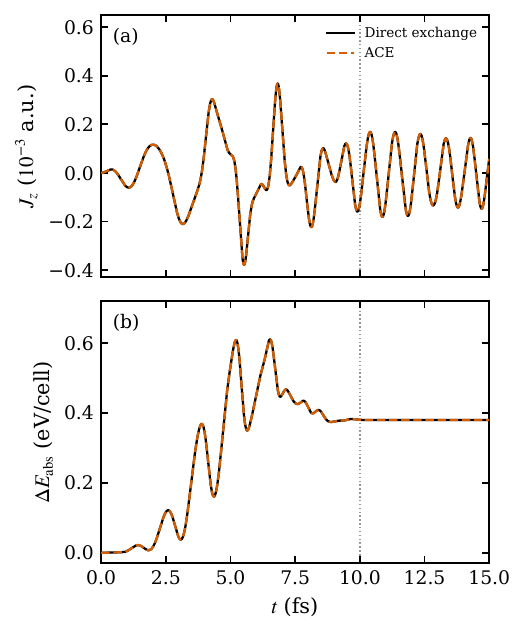}
\caption{Comparison of direct screened exchange and ACE without temporal prediction for (a) the charge-current density and (b) the absorbed energy per Si cell. The laser photon energy is 1.55 eV, the peak intensity is $10^{12}\,\mathrm{W\,cm^{-2}}$, and the pulse has 10-fs full support. The dotted line marks the pulse end.}
\label{fig:ace-direct-waveforms}
\end{figure}

\section{Comparison of ACE and direct exchange}
\label{app:ace-direct}
To check the accuracy of ACE before introducing temporal prediction, we compare it with direct screened exchange under the laser-driven conditions of Sec.~\ref{sec:learned-validation}. Both calculations use identical initial orbitals, grid, $\bm k$ mesh, field, and time integrator.

\begin{figure}[!t]
\centering
\includegraphics[width=\columnwidth]{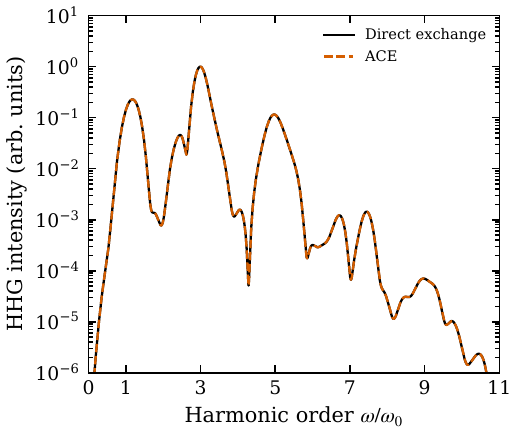}
\caption{HHG spectra for direct screened exchange and ACE without temporal prediction. Both use the same 15-fs window and are normalized by the maximum direct-exchange intensity over the displayed range.}
\label{fig:ace-direct-hhg}
\end{figure}

Figure~\ref{fig:ace-direct-waveforms} compares the current and absorbed energy. The full-record relative current difference is $0.15\%$. The final absorbed-energy difference is $0.010\%$. Figure~\ref{fig:ace-direct-hhg} shows that the HHG spectra also closely agree, using the same 15-fs window and spectral definition in Sec.~\ref{sec:measures}.

\section{Comparison with linear extrapolation}
\label{app:linear-prediction}

The simplest predictor extends the change between the two latest explicit factors,
\begin{equation}
\widehat{\Xi}_{m,\ell}^{\mathrm{lin}}=\Xi_m+s(\Xi_m-\Xi_{m-1}),
\end{equation}
using the endpoint and interval notation of Sec.~\ref{sec:learned-method}.
This predictor retains gauge alignment and periodic explicit corrections but requires no fitted coefficients.

To determine whether fitting the coefficients improves prediction, we compare linear extrapolation and learned ACE against conventional ACE under identical conditions (Sec.~\ref{sec:learned-validation}).

\begin{table}[!t]
\caption{Errors of linear extrapolation and learned prediction at $K=8$. The current error is integrated over the full trajectory, and the absorbed-energy error is the relative difference at the final time.}
\begin{ruledtabular}
\begin{tabular}{lrr}
Predictor & Current error (\%) & Energy error (\%)\\
\hline
Linear extrapolation & 10.68 & 0.55\\
Learned prediction & 3.12 & 0.90\\
\end{tabular}
\end{ruledtabular}
\end{table}
For the tested correction intervals $K=2$, 4, 8, and 12, linear extrapolation gives current errors approximately 2.5--3.4 times those of learned prediction. At $K=8$, both final absorbed-energy errors are below 1\%, with linear extrapolation giving the smaller error. The ratio of RT computation times for linear extrapolation to learned prediction is approximately 0.99. Learned prediction therefore improves current fidelity at essentially the same cost.

\section{Choice of time propagator}
\label{app:time-propagator}
Parallel-transport implicit propagators allow larger time steps but require iterative solution of the time-dependent equations \cite{Jia2019,Liu2025}. In preliminary tests of our implementation, repeated ACE reconstruction and Hamiltonian applications within these iterations offset the benefit of the larger time steps. We therefore use fourth-order Taylor propagation with predictor--corrector updates. Learned ACE reduces reconstruction frequency directly: prediction steps require only factor combinations with two fitted coefficients, without an iterative exchange solve. Gauge alignment is performed only at explicit history updates.

\bibliographystyle{apsrev4-2}
\bibliography{references,bib-control}
\end{document}